\documentclass[sigconf]{acmart}
\usepackage[utf8]{inputenc} 
\usepackage{xcolor}
\usepackage{pifont}
\usepackage{booktabs}
\usepackage{graphicx}
\usepackage{tcolorbox}
\usepackage{multirow}
\usepackage{enumitem}
\usepackage{colortbl}
\usepackage{subcaption} 
\usepackage{todonotes}
\usepackage{xspace}
\usepackage{threeparttable}
\usepackage{listings}
\usepackage{cleveref}

\AtBeginDocument{%
  }

\setcopyright{acmlicensed}
\copyrightyear{2025}
\acmYear{2025}
\acmDOI{XXXXXXX.XXXXXXX}

\definecolor{mygreen}{rgb}{0,0.6,0}
\definecolor{myred}{rgb}{1,0,0}
\definecolor{diffgreen}{rgb}{0.85,0.95,0.85}
\definecolor{diffred}{rgb}{1,0.85,0.85}
\definecolor{diffgreentext}{rgb}{0,0.5,0}
\definecolor{diffredtext}{rgb}{0.7,0,0}
\definecolor{myblue}{rgb}{0,0,0.8}
\definecolor{lightgray}{rgb}{0.95,0.95,0.95}

\usepackage{listings}
\crefname{lstlisting}{listing}{listings}
\Crefname{lstlisting}{Listing}{Listings}
\usepackage{xcolor}

\definecolor{codegreen}{rgb}{0,0.6,0}
\definecolor{codegray}{rgb}{0.5,0.5,0.5}
\definecolor{codepurple}{rgb}{0.58,0,0.82}
\definecolor{shallowred}{rgb}{1,0.8,0.8}
\definecolor{verylightgray}{rgb}{0.97, 0.97,0.97}

\lstdefinestyle{mystyle}{
    language=C++, 
    commentstyle=\color{codegreen},
    keywordstyle=\color{blue},
    stringstyle=\color{codepurple},
basicstyle=\fontfamily{zi4}\selectfont\scriptsize,
    breakatwhitespace=false,
    breaklines=true,              
    captionpos=b,                    
    keepspaces=true,                 
    numbers=left,                    
    numbersep=5pt,                  
    showspaces=false,                
    showstringspaces=false,
    frame=single,                    
    xleftmargin=0.05\linewidth,
    xrightmargin=0.05\linewidth,
    framexleftmargin=0.05mm,            
    framexrightmargin=0.05mm,           
    framextopmargin=0mm,             
    framexbottommargin=0mm,    
    showtabs=false,                  
    tabsize=2,
    escapeinside={(*@}{@*)},
}

\acmConference[Conference acronym 'XX]{Make sure to enter the correct
  conference title from your rights confirmation email}{Feb 24,
  2025}{Woodstock, NY}

\newcommand{\find}[1]{
\begin{tcolorbox}[leftrule=1mm,toprule=0mm,bottomrule=0mm,left=1pt,right=2pt,top=2pt,bottom=2pt]
\em #1
\end{tcolorbox}
}

\newcommand{\method}{{\textsc{PatchSeeker}}\xspace}
\begin{document}

\setcounter{page}{1}
\title[\textsc{\method}]{\method: Mapping NVD Records to their Vulnerability-fixing Commits with LLM Generated Commits and Embeddings}

\author{Huu Hung Nguyen}
\affiliation{%
  \institution{Singapore Management University}
  \country{Singapore}
}
\email{huuhungn@smu.edu.sg}

\author{Anh Tuan Nguyen}
\email{tuan.na236009@sis.hust.edu.vn}
\affiliation{%
  \institution{Hanoi University of Science and Technology}
  \country{Vietnam}
}

\author{Thanh Le-Cong}
\email{congthanh.le@student.unimelb.edu.au}
\affiliation{%
  \institution{University of Melbourne}
  \country{Australia}
}

\author{Yikun Li}
\affiliation{%
  \institution{Singapore Management University}
  \country{Singapore}
}
\email{yikunli@smu.edu.sg}

\author{Han Wei ANG}
\affiliation{%
  \institution{GovTech}
  \country{Singapore}
}
\email{ANG_Han_Wei@tech.gov.sg}

\author{Yide Yin}
\affiliation{%
  \institution{GovTech}
  \country{Singapore}
}
\email{YIN_Yide@tech.gov.sg}

\author{Frank Liauw}
\affiliation{%
  \institution{GovTech}
  \country{Singapore}
}
\email{Frank_LIAUW@tech.gov.sg}

\author{Shar Lwin Khin}
\affiliation{
\institution{Singapore Management University}
\country{Singapore}
}
\email{lkshar@smu.edu.sg}

\author{Ouh Eng Lieh}
\affiliation{
\institution{Singapore Management University}
\country{Singapore}
}
\email{elouh@smu.edu.sg}

\author{Ting Zhang}
\affiliation{%
 \institution{Monash University}
 \country{Australia}}
\email{ting.zhang@monash.edu}
\authornote{Ting Zhang is the corresponding author.}

\author{David Lo}
\affiliation{%
  \institution{Singapore Management University}
  \country{Singapore}
}
\email{davidlo@smu.edu.sg}

\renewcommand{\shortauthors}{Nguyen et al.}

\begin{abstract}

Software vulnerabilities pose serious risks to modern software ecosystems.
While the National Vulnerability Database (NVD) is the authoritative source for cataloging these vulnerabilities, it often lacks explicit links to the corresponding Vulnerability-Fixing Commits (VFCs).
VFCs encode precise code changes, enabling vulnerability localization, patch analysis, and dataset construction.
Automatically mapping NVD records to their true VFCs is therefore critical. 
Existing approaches have limitations as they rely on sparse, often noisy commit messages and fail to capture the deep semantics in the vulnerability descriptions.

To address this gap, we introduce \method, a novel method that leverages large language models to create rich semantic links between vulnerability descriptions and their VFCs.
\method generates embeddings from NVD descriptions and enhances commit messages by synthesizing detailed summaries for those that are short or uninformative. 
These generated messages act as a semantic bridge, effectively closing the information gap between natural language reports and low-level code changes.

Our approach \method achieves 59.3\% higher MRR and 27.9\% higher Recall@10 than the best-performing baseline, Prospector, on the benchmark dataset.
The extended evaluation on recent CVEs further confirms \method's effectiveness.
Ablation study shows that both the commit message generation method and the selection of backbone LLMs make a positive contribution to \method.
We also discuss limitations and open challenges to guide future work.
\end{abstract}
\maketitle


\section{Introduction}

With the growth of the software ecosystem, the number of software vulnerabilities also increases rapidly. 
Last year, NIST's National Vulnerability Database (NVD) recorded 40,009 new CVE entries with an average of 108 new disclosures each day~\cite{nvd}. 
This surge necessitates the effective management of known vulnerabilities to identify and remediate their impact on software systems. 
To fulfill this demand, security platforms, such as the NVD, have been established to catalog the detailed information about disclosed vulnerabilities, such as vulnerability descriptions, CVSS severity rating, and links to vulnerability-fixing commits (VFCs).
Among them, VFCs have recently attracted much research interest as they contain raw information for both the vulnerability and its fix.
Thus, VFCs can be used for many downstream tasks, such as vulnerable software version identification~\cite{cheng2024llm}, vulnerability detection~\cite{zhou2024large}, and vulnerability repair~\cite{zhou2024out}.

Yet, VFC links are largely missing from the NVD.
A recent empirical study~\cite{nguyen2025mappingnvdrecordsvfcs} found that approximately 93.3\% of the CVEs lack explicit links in NVD to their corresponding VFCs.
This absence of VFCs is a major obstacle for the security community. 
Without a sizeable, accurately linked corpus of VFCs, learning‑based techniques cannot reach their full potential in vulnerability discovery or patch generation.
Automatically mapping NVD records\footnote{Note: As each NVD record describes a unique CVE, we use the terms "NVD record" and "CVE" interchangeably. Similarly, "CVE description" refers to the description of such vulnerability in a NVD record.} to their VFCs has therefore become challenging for software security~\cite{prospector,bhandari2021cvefixes}.

Recent approaches have tried to solve this mapping problem using either heuristic or learning-based methods. 
Heuristic tools such as Prospector~\cite{prospector} rely on fixed rules and keyword matching. 
Learning-based tools, most notably PatchFinder~\cite{li2024patchfinder}, use a two-phase approach to filter and rank commits. 
However, our evaluation shows that further improvement over these existing methods is needed, as the top-ranked result from Prospector and PatchFinder only achieved Recall@1 at 0.354 and 0.2, respectively, on our rigorously built benchmark dataset.

Our analysis reveals that these existing methods suffer from two key limitations (see details in Section~\ref{sec:motivating_examples}):

\vspace{2px}
\noindent \textbf{Limitation I - Poor semantic understanding in CVE descriptions and commit messages.} Existing learning-based methods~\cite{li2024patchfinder, patchscout} rarely capture the deeper semantic link between a CVE description and the commit message of its VFC.
Even when a commit message contains the same keywords mentioned in the CVE description partially, existing learning-based models often fail to make the connection.
Meanwhile, heuristic-based techniques~\cite{prospector} rely on a set of predefined rules to map CVE descriptions to the commit messages of their VFCs.
Although these rules can capture relationships between NVD descriptions and commit messages more precisely, they are constrained by human efforts in designing these rules, leading to limited flexibility.

\vspace{2px}
\noindent \textbf{Limitation II - Uninformative commit messages.} These methods rely too heavily on commit messages, while many vulnerabilities are fixed ``silently"~\cite{cheng2025fixseeker, zhou2021finding, nguyen2023multi}. 
It is a common practice by software developers to use vague commit messages without explicitly mentioning security-related information or internal ticket numbers to avoid drawing attention from attackers to exploit vulnerabilities~\cite{zhou2023colefunda,zhou2021finding}. 
This leaves the models with no meaningful information from the commit message to analyze, making it nearly impossible to find the correct VFC.

To address these limitations, this paper introduces \method that utilizes automated commit message generation and embedding from Large Language Models (LLMs) to better model the similarity between CVE descriptions and the commit messages of their VFCs.
Specifically, to address \textbf{Limitation I}, we leverage the semantic understanding capability of modern LLMs to tackle this complex mapping task. 
These LLMs are trained on extensive textual datasets, allowing them to capture semantic relationships between texts more accurately than traditional and smaller-scale language models used in prior works.
To address \textbf{Limitation II}, we leverage CCT5~\cite{lin2023cct5}, a state-of-the-art code-change-oriented language model, to generate more informative commit messages. 
This approach allows \method to increase the chance of matching the CVE description with the commit messages, which originally contain limited information.

To evaluate the effectiveness of \method, we rigorously built a new dataset of 5,000 CVEs by matching NVD entries with their VFCs from the Morefixes dataset~\cite{morefixes}, prioritizing high-confidence pairs and focusing on the top-25 most dangerous CWEs~\cite{mitre_top25} across 2,070 repositories, with temporal constraints ensuring realistic negative sampling. We compare \method against three state-of-the-art baselines: PatchFinder~\cite{li2024patchfinder} and PatchScout~\cite{patchscout} (learning-based approaches) and Prospector~\cite{prospector} (a heuristic-based tool). Our experiments show that \method achieves a Recall@10 of 0.871 and MRR of 0.739, which substantially outperforms the best-performing baseline by 27.9\% and 59.3\%, respectively. 
Moreover, we also demonstrate the real-world applicability of our approach by evaluating \method on a set of recent CVEs. 
This set includes both CVEs with explicit links to VFCs and those lacking associated VFCs.
On CVEs with explicit links to VFCs, our approach achieves a Recall@10 of 0.786, which is 27\% higher than the best-performing baseline. 
More importantly, on the more challenging CVEs without explicitly associated VFCs, \method demonstrates a Precision@10 of 0.710, representing a 97\% improvement over the best-performing baseline. 

Our contributions can be summarized as follows:
\begin{itemize}[leftmargin=*]
    \item We present \method, a novel approach that couples LLM-based semantic encoding and commit message generation for mapping NVD records to their VFCs.
    \item We construct and release the \method dataset, which consists of 5,000 CVEs across 2,070 repositories. 
    The dataset focuses on the top 25 most dangerous CWEs. 
    \item Our experimental results show that \method significantly outperforms the baselines on our benchmark, achieving a 27.9\% improvement in Recall@10, a 59.3\% improvement in MRR over the best-performing baseline. It also reduces the Manul Efforst@10 by 50.8\% than the best-performing baseline.
\end{itemize}

The remaining part of this paper is organized as follows: Section~\ref{sec:background} includes motivating examples and problem formulation.
Section~\ref{sec:approach} elaborates on our proposed approach \method.
In Section~\ref{sec:setup}, we provide details about the experimental setup.
Experimental results are presented in Section~\ref{sec:results}.
We also present qualitative analysis in Section~\ref{sec:discussion}.
We discuss related work in Section~\ref{sec:related}.
Section~\ref{sec:conclusion} concludes and discusses the future work.

\section{Background}
\label{sec:background}

\subsection{Motivating Examples}
\label{sec:motivating_examples}


\begin{lstlisting}[caption=Example of Poor Semantic Understanding from PatchFinder,label={lst:motivating_1},float=t]
(1) CVE Description of CVE-2017-12987:
"The IEEE 802.11 parser in tcpdump before 4.9.2 has a buffer over-read in print-802_11.c:parse_elements()."

(2) Commit message of the true VFC. Discarded by Phase 1 of PatchFinder: 
"CVE-2017-12987/IEEE 802.11: Fix processing of TIM IE. The arguments to memcpy() were completely wrong. This fixes a buffer over-read discovered by Kamil."

(3) Commit message of a false VFC that passed Phase 1 of PatchFinder: 
"RESP: Remove some redundant checks. Before we break out of the loop, we've already checked for those conditions. No need to check for them again. This fixes Coverity CIDs 1400553 and 1400554."
\end{lstlisting}

\noindent\textbf{Challenge 1: Poor semantic understanding in NVD description and VFC commit message.}
As the prior state-of-the-art learning-based method, PatchFinder~\cite{li2024patchfinder} employs a two-phase approach that relies on a hybrid of lexical matching, TF-IDF~\cite{tfidf}, and a small-sized pre-trained language model, CodeReviewer~\cite{li2022codereviewer}. 
The core problem is that TF-IDF is limited to surface-level keyword matching, and CodeReviewer lacks the deep contextual reasoning of modern LLMs. 
This leads to a critical issue: the initial filtering phase can prematurely eliminate the true VFC before it can be further analyzed in the second phase.
On our \method dataset, we found that for 1,989 out of 5,000 CVEs (39.78\%), the true VFC was discarded in PatchFinder's Phase 1.

Listing~\ref{lst:motivating_1} shows a clear example of this, which is from CVE-2017-12987, representing a buffer over-read vulnerability in the TCPdump project~\footnote{https://github.com/the-tcpdump-group/tcpdump}. The true VFC's commit message explicitly mentioned both the CVE ID and "buffer over-read." However, the shallow lexical analysis of TF-IDF and the limited semantic understanding of CodeReviewer failed to assign it a high enough score, causing it to be incorrectly discarded. This failure to connect obvious semantic links highlights that these older components are simply not enough for this task.

Furthermore, even when a commit survives Phase 1, PatchFinder's design remains suboptimal. 
By concatenating the commit message with the raw code diff, the length of the input becomes larger, thus the model's attention on the semantically rich commit message gets weaker. 
Additionally, its use of a pointwise Binary Cross-Entropy loss function is not ideal for a ranking task, where a pairwise contrastive loss function is more effective for retrieval tasks~\cite{wan2022cross, jin2023debiased}. 
These limitations highlight the need for a new approach, which can leverage the reasoning and holistic understanding capabilities of larger-sized LLMs.


\begin{lstlisting}[caption=Example to illustrate the correlations between a vulnerability and its VFC.,label={lst:motivating_2},float=t]
(1) CVE Description of CVE-2023-27898:
"Jenkins 2.270 through 2.393 (both inclusive), LTS 2.277.1 through 2.375.3 (both inclusive) does not escape the Jenkins version a plugin depends on when rendering the error message stating its incompatibility with the current version of Jenkins, resulting in a stored cross-site scripting (XSS) vulnerability exploitable by attackers able to provide plugins to the configured update sites and have this message shown by Jenkins instances."

(2) VFC of CVE-2023-27898
Commit message: "[SECURITY-3037]"

Code diff:
diff -- git core/src/main/java/hudson/PluginManager.java
@@ -1469,7 +1469,11 @@ public HttpResponse doPluginSearch(...)
     jsonObject.put("sourceId", plugin.sourceId);
     jsonObject.put("title", plugin.title);
     jsonObject.put("displayName", plugin.getDisplayName());
(*@\colorbox{diffred}{\makebox[\linewidth][l]{\textcolor{diffredtext}{  -     jsonObject.put("wiki", plugin.wiki);}}}@*)
(*@\colorbox{diffgreen}{\makebox[\linewidth][l]{\textcolor{diffgreentext}{  +     if (plugin.wiki == null || !(plugin.wiki.}}}@*)
(*@\colorbox{diffgreen}{\makebox[\linewidth][l]{\textcolor{diffgreentext}{         startsWith("https://") ||}}}@*)
(*@\colorbox{diffgreen}{\makebox[\linewidth][l]{\textcolor{diffgreentext}{  +    plugin.wiki.startsWith("http://"))) \{ }}}@*)
(*@\colorbox{diffgreen}{\makebox[\linewidth][l]{\textcolor{diffgreentext}{  +         jsonObject.put("wiki", StringUtils.EMPTY);}}}@*)
(*@\colorbox{diffgreen}{\makebox[\linewidth][l]{\textcolor{diffgreentext}{ +     \} else \{}}}@*)
(*@\colorbox{diffgreen}{\makebox[\linewidth][l]{\textcolor{diffgreentext}{ +         jsonObject.put("wiki", plugin.wiki);}}}@*)
(*@\colorbox{diffgreen}{\makebox[\linewidth][l]{\textcolor{diffgreentext}{ +     \}}}}@*)
       jsonObject.put("categories", plugin.
              getCategoriesStream()
...
\end{lstlisting}


\vspace{2px}
\noindent\textbf{Challenge 2: Uninformative commit messages.}
A second major challenge comes from the commit messages themselves. In many cases, the lack of detail is not accidental but is an intentional security measure. 
Following recommended practices like Coordinated Vulnerability Disclosure (CVD)~\cite{cvd}, developers often avoid mentioning vulnerability-related keywords in public commit messages to prevent attackers from easily discovering and exploiting the flaws before users can apply the patch~\cite{zhou2023colefunda}. 
These are known as ``silent vulnerability fixes"~\cite{cheng2025fixseeker}, which is common in open-source software (OSS) management. 
In fact, studies show that one in four open-source projects fix security bugs silently, without ever announcing the fix publicly~\cite{sawadogo2020learning,nguyen2023vffinder}. 
While this responsible strategy protects users, it creates a significant obstacle for automated VFC linking tools. 
As the commit message is deliberately disconnected from the vulnerability it fixes, it leaves models with no information to analyze.

Listing~\ref{lst:motivating_2} demonstrates a clear example of this problem, which involves CVE-2023-27898, an XSS bug in Jenkins. 
The official commit that fixes this bug (\texttt{59ac866d...}) has a not very helpful message: \texttt{[SECURITY-3037]}. 
This message is unclear without access to the Jenkins issue tracker and offers no keywords or descriptive text for an automated tool to use.
However, the code change itself is much more informative. 
To summarize, the code clearly shows an XML escaping function (\texttt{Util.xmlEscape}) being added to the \texttt{PluginManager.java} file, which is a standard way to prevent XSS. 
This leads to a key observation: the code diff explains the fix but the message does not.

This is why a commit message generation model is important, and how the information from the code diff is used matters significantly. 
As mentioned in Challenge 1, while a tool like PatchFinder also incorporates the code diff, it does so by concatenating the raw code directly with the message. 
This approach can be suboptimal, as the noisy syntax of a raw diff can distract attention from key semantic signals. 
Instead of using the raw code, our approach generates a concise, natural-language summary from it. 
For instance, by looking at the informative code change, a model like CCT5~\cite{lin2023cct5} could create a much better summary, such as: ``Fix XSS vulnerability in PluginManager by sanitizing plugin metadata.'' 
By adding this richer, generated message to the original short one, a model can overcome the problem of poor commit messages and use the helpful information found directly in the code. 

\subsection{Problem Formulation}
Given a CVE description $v$, our task is to identify its VFCs from a set of candidate commits.
Formally, for each CVE $v$, we have a set of candidate commits $\mathcal{C} = \{c_1, c_2, \ldots, c_n\}$. The details of collecting these candidates can be found in Section~\ref{sec:dataset}.
Our goal is to compute a score vector ${S} = \{s_1, s_2, \ldots, s_n\}$, where each score $s_i \in [0, 1]$ represents the probability that candidate $c_i$ is the true VFC. 
This scoring mechanism effectively re-ranks the candidates in $\mathcal{C}$ such that true VFCs receive the highest scores.

\section{Approach}
\label{sec:approach}


\begin{figure*}[h]
\centering
\includegraphics[width=0.8\textwidth]{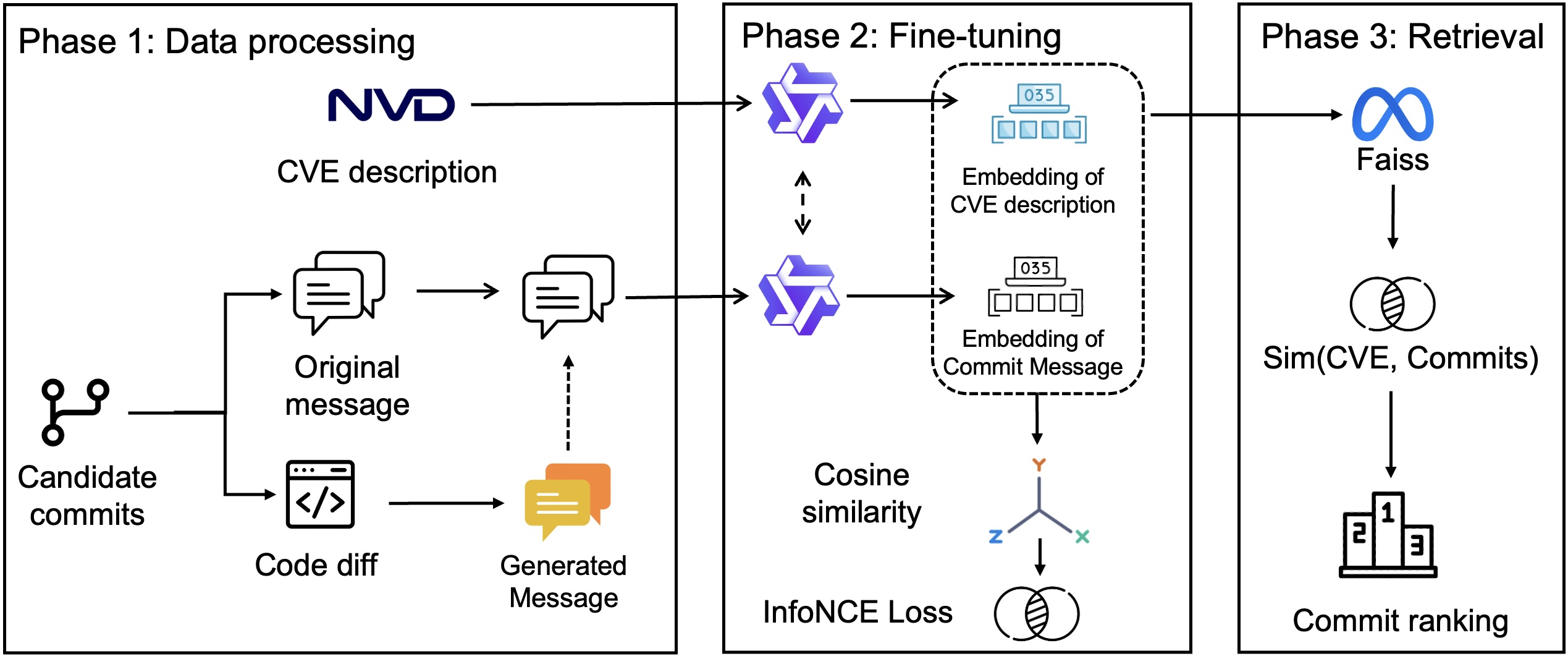} 
\caption{\method is a three-phase system for identifying VFCs. In Phase 1, it processes CVE descriptions and their corresponding commit messages. In Phase 2, it fine-tunes an LLM using a bi-encoder architecture with an InfoNCE loss function. In Phase 3, this fine-tuned LLM is used as an encoder to retrieve and rank candidate VFCs.}
\label{fig:overview}
\end{figure*}


\subsection{Overview}
As illustrated in Figure~\ref{fig:overview}, our proposed approach~\method consists of three main phases as follows.

\begin{itemize}[leftmargin=*]
\item Data Processing: For a given CVE description and a set of candidate commits, this stage prepares the necessary inputs. 
It extracts the developer-written commit message and, optionally, uses a commit message model to generate an additional, more descriptive message from the commit diff. 
These messages are then combined to form a richer input for the next phase.

\item Fine-tuning: The goal of this stage is to fine-tune an LLM to understand the semantic relationship between a CVE description and its true VFC. We use Qwen3 as our base model and employ a contrastive learning objective (InfoNCE loss~\cite{infonce}) to teach the model to assign higher similarity scores to the CVE description and the commit message of its true VFCs compared to the commit message of its non-VFCs.

\item Retrieval: In this phase, the fine-tuned LLM is used as an offline encoder. 
For a new CVE description, we generate embeddings for it and for all its candidate commits. 
We then use Faiss~\cite{douze2024faiss}, a library for efficient similarity search, to rapidly retrieve and rank the candidate commits based on their proximity to the NVD description, producing a final ranked list of potential VFCs.
\end{itemize}


\subsection{Phase 1: Data Processing}
The first phase of our approach focuses on preparing the input data from the developer-written commit messages and their associated code diffs.

\subsubsection{Commit Message Augmentation with CCT5}
While the primary \method workflow relies on the semantic content of developer-written commit messages, these messages can be too brief or lack sufficient detail to establish a strong connection with a CVE description. 
To address this limitation and leverage the rich information contained within the commit diffs, we introduce an additional technique using the CCT5 model~\cite{lin2023cct5} for automated commit message generation. 
We select the CCT5 model as it is the state-of-the-art approach for commit message generation~\cite{fan2024exploring}.
This approach serves as a parallel branch to enrich the input data, as illustrated by the CCT5 path in Figure~\ref{fig:overview}.

For the commits that are too brief, we follow the methodology of CCT5, providing the commit diff as input to generate a new, descriptive commit message. 
To handle practical constraints and align with the model's architecture, we follow the original setting of CCT5, which is truncating the input commit diffs to a maximum length of 100 tokens.
The newly generated message is then concatenated with the original developer's message. 
This creates a single, more descriptive text input that is subsequently fed into the LLM for embedding. 


\subsection{Phase 2: Fine-tuning}
The training phase aims to fine-tune an LLM to act as a specialized encoder that maps semantically related CVE descriptions and VFC messages to nearby points in an embedding space. 
We adopt a bi-encoder dense retriever architecture, inspired by prior studies~\cite{repllama,karpukhin2020dense}.

\subsubsection{Bi-Encoder Architecture}
To capture the deep semantic meaning of the text, we use the Qwen~3~\cite{qwen3} as our encoder. 
Both the NVD description text $d(v)$ and each candidate commit message $m(c_i)$ (where $c_i \in \mathcal{C}$) are passed through the model. 
We then extract the last hidden state vector of the final token as the embedding representation for each piece of text. 
This process yields a dense vector embedding ${e}_d \in \mathbb{R}^d$ for the NVD description and a set of embeddings $\{{e}_{c_i} \in \mathbb{R}^d\}$ for each candidate commit. 
These embeddings serve as the numerical representation of the text for the subsequent similarity computation.

\subsubsection{Contrastive Learning with InfoNCE Loss}
To adapt the LLM for our retrieval task, we follow the prior study~\cite{repllama} to employ the InfoNCE loss function~\cite{infonce}, which is widely used in contrastive learning for learning discriminative representations~\cite{shao2021temporal, wu2021rethinking}. 
Given a CVE description with its true VFCs as the positive samples and its non-VFCs as negative samples, the InfoNCE loss encourages the model to maximize the similarity between the CVE description and the commit messages of its true VFCs while minimizing similarities with the commit messages of its non-VFCs.


Formally, for a CVE $v$ with its CVE description embedding ${e}_d$ and a set of candidate commit embeddings $\{{e}_{c_1}, {e}_{c_2}, \ldots, {e}_{c_n}\}$ where $c_k$ is the true VFCs, the InfoNCE loss is defined as:

\begin{equation}
\mathcal{L}_{\text{InfoNCE}} = -\log \frac{\exp(\text{sim}({e}_d, {e}_{c_k}) / \tau)}{\sum_{i=1}^{n} \exp(\text{sim}({e}_d, {e}_{c_i}) / \tau)}
\label{eq:infonce}
\end{equation}

Here, $\text{sim}(\cdot, \cdot)$ is the cosine similarity function, used during training to provide an exact similarity score for backpropagation.
$\tau > 0$ is a temperature hyperparameter that controls the sharpness of the distribution.
$c_k$ represents the true VFCs among all candidates. 
The numerator computes the exponential of the scaled similarity between the CVE description and the true VFCs, while the denominator sums over all candidates, including both positive and negative samples.
By minimizing this loss, the model learns to organize its embedding space such that true pairs are pulled together and incorrect pairs are pushed apart.

\subsection{Phase 3: Retrieval}

In the retrieval phase, we use the trained model to efficiently identify and rank potential VFCs for a new CVE description.
First, the fine-tuned Qwen-3 model is used as an encoder to generate an embedding for the new NVD description and embeddings for the entire set of candidate commits.

Next, to quickly compare a single CVE description against thousands of candidate commits, we use Faiss~\cite{douze2024faiss} for fast similarity search in large-scale vector datasets. 
Faiss first indexes the embeddings of all candidate commits.
Then, given the embedding of a new CVE description, Faiss rapidly retrieves the top-k most similar commit embeddings from the index, yielding the final ranked list of candidate VFCs. 
This allows \method to scale effectively to real-world scenarios with extensive historical commits.

\section{Experimental Setup}
\label{sec:setup}
In this section, we first introduce the research questions (RQs) we plan to answer in this work.
Second, we mention how we construct the dataset.
Third, we briefly describe the state-of-the-art VFC identification approaches.
Next, we discuss the evaluation metrics adopted in this study.
Finally, we provide implementation details.

\subsection{Research Questions}
\textbf{RQ1: How does \method perform compared to the baselines?}
\noindent\textit{Motivation}: To demonstrate the effectiveness of \method, we benchmark it against both learning‑based and heuristic state‑of‑the‑art approaches for VFC identification. 
Evaluating across this diverse set of baselines is crucial for showing that \method delivers consistent, significant gains in accurately mapping NVD records to their VFCs.

\noindent\textit{Approach}: We conduct a comprehensive evaluation by comparing \method against three baseline methods: learning-based models, i.e., PatchScout~\cite{patchscout} and PatchFinder~\cite{li2024patchfinder}, and a heuristic-based tool, i.e., Prospector~\cite{prospector}. 

\vspace{2px}
\noindent\textbf{RQ2: How does each component of \method contribute to its overall performance?}

\noindent\textit{Motivation}: Understanding the individual contributions of \method's components is crucial for explaining its superior performance and guiding future improvements. 
Specifically, we investigate two key design choices: the integration of CCT5 for generating commit messages and the selection of backbone LLMs that serve as the core embedding engine. 

\noindent\textit{Approach}: We conduct an ablation study examining two critical components of \method. 
First, to evaluate CCT5's contribution, we compare \method's performance with and without CCT5-generated commit messages using the same backbone model of Qwen 3, measuring the impact on all three evaluation metrics.
Second, to assess the influence of backbone model selection, we have replaced the Qwen 3 with other LLMs.
Specifically, we replace the Qwen 3 (\texttt{Qwen3-8B}) used in the final \method with three different LLMs as embedding generators: CodeReviewer~\cite{li2022codereviewer}, LLaMA 2~\cite{llama2} (\texttt{Llama-2-7b}), LLaMA 3.2~\cite{llama3} (\texttt{Meta-Llama-3-8B}).

\vspace{2px}
\noindent\textbf{RQ3: What is the performance of \method on recent CVEs?} \textit{Motivation}: To demonstrate that \texttt{\method} can handle evolving vulnerability patterns and is not overfitting to historical data, we conduct an additional evaluation on CVEs published within the last six months. 
We categorize these CVEs based on whether their VFCs are explicitly referenced in the NVD. 
Some CVEs feature \textit{explicit} VFCs, which are identifiable by Git-based links labeled as \texttt{Patch} in their NVD reference section.
However, the majority have \textit{implicit} VFCs, lacking any direct VFC links. 
This latter group is highly significant, as a recent study found they constitute 93.4\% of all CVEs in the NVD~\cite{nguyen2025mappingnvdrecordsvfcs}. 
This prevalence underscores the practical need for an automated approach capable of mapping NVD records to their implicit VFCs, a critical task for real-world security.

\vspace{2px}
\noindent\textit{Approach}: We first crawled all the NVD records published between December 22, 2024, and June 22, 2025, dividing them into (1) \texttt{explicit-set}, which contains 84 CVEs with explicit VFC links in NVD, and (2) \texttt{implicit-set}, which contains 100 randomly sampled CVEs with implicit VFCs.
Second, we evaluate \method and the two best-performing baselines based on the RQ2 experimental results, i.e., PatchFinder and Prospector.
For CVEs in \texttt{explicit-set}, we compute standard metrics (Recall@K, MRR, and Manual Effort@K, which are widely used in prior VFC linking work~\cite{patchscout,li2024patchfinder,wang2022vcmatch}) using NVD-provided ground truth VFCs. 
For CVEs in \texttt{implicit-set}, two of the authors manually validate the top-10 ranked candidates produced by PatchFinder, Prospector and \method to determine whether such a candidate is a true VFC.
We then report Manual Effort@K and Precision@K metrics, where K is capped at 10.

\subsection{Dataset Construction}
\label{sec:dataset}
We construct the main benchmark dataset by collecting CVEs from the NVD and matching them with their corresponding VFCs from the Morefixes dataset~\cite{morefixes}. 
To ensure the quality of the ground-truth data, we leverage the confidence score provided by Morefixes for each CVE-VFC pair and prioritize those with the highest scores, which typically correspond to VFCs explicitly identified by the NVD.
To focus on the most critical security vulnerabilities, we filter the dataset to include only CVEs associated with the top-25 most dangerous CWEs in 2024 according to MITRE~\cite{mitre_top25} 
resulting in 5,000 CVEs spanning 2,070 repositories. 
We randomly split these CVEs into training (80\%), validation (10\%), and test (10\%) sets to ensure robust evaluation.


For negative sample selection, we adopt a temporal constraint strategy based on empirical analysis. 
Our examination of the PatchFinder dataset revealed that 91.2\% of VFCs occur within one year before the CVE's NVD publication date, indicating that vulnerability fixes often occur less than a year before their public disclosure.
Therefore, for each CVE, we include all the candidate commits from the corresponding repository within the one-year window before the NVD publication date as candidate commits, averaging 6,621 negative samples per CVE. 

We choose to construct our dataset rather than using the existing PatchFinder dataset for several key reasons. 
First, our focus on the top-25 most dangerous CWEs ensures evaluation on high-impact vulnerabilities, while PatchFinder includes CVEs without CWE-based prioritization.
Second, our dataset provides substantially broader repository coverage (2,070 vs. 510 repositories), better representing the diversity of real-world software projects.
Third, our temporal constraint for negative sampling creates a more realistic and challenging evaluation scenario compared to PatchFinder's random sampling approach, which may include commits far removed from the vulnerability timeline. 
Finally, PatchFinder only provides CVE-VFC pairs without negative samples or dataset splits, limiting reproducibility and fair comparison across methods. 
Instead, we share the URLs for the negative samples to ensure the future work can evaluate on the same set of data.
\subsection{Baselines}
We compare \method against three state-of-the-art approaches, i.e., PatchScout~\cite{patchscout} and PatchFinder~\cite{li2024patchfinder} are learning-based, and Prospector~\cite{prospector} is heuristic-based.
\begin{itemize}[leftmargin=*]
    \item \textbf{PatchScout~\cite{patchscout}} leverages vulnerability-commit correlation ranking through a three-phase workflow: extracting information elements from NVD descriptions and commits, generating correlation features between vulnerabilities and commits, and training a RankNet-based model~\cite{burges2010ranknet} to rank commits by their relevance to vulnerabilities. The method focuses on learning statistical patterns that correlate vulnerability characteristics with commit properties.
    \item \textbf{PatchFinder~\cite{li2024patchfinder}} employs a two-phase framework: (1) an initial retrieval phase using TF-IDF and a CodeReviewer for lexical and semantic similarity scoring, (2) a re-ranking phase fine-tuning the CodeReviewer to capture precise code semantics for security patches. This approach effectively filters out unrelated commits in the first phase before applying deep semantic understanding to identify true VFCs among top candidates in the second phase.
    \item \textbf{Prospector~\cite{prospector}} is a rule-based tool that extracts keywords from NVD descriptions using NLP techniques, identifies security-relevant keywords and file/class/method names through pattern matching, and applies a set of heuristic rules to rank candidate commits. This approach relies on manually crafted rules and keyword matching rather than learning from data, making it interpretable but potentially less adaptable to diverse vulnerability patterns.
\end{itemize}

\subsection{Evaluation Metrics}
Following recent studies on VFC retrieval~\cite{li2024patchfinder,patchscout,wang2022vcmatch}, we evaluate VFC identification performance using three complementary ranking metrics that capture different aspects of VFC retrieval quality: Recall@K, MRR, and Manual Effort@K.
Recall@K and MRR are typical information-retrieval metrics, while Manual Efforts@K emerges as a classic metric in tracing VFCs within OSS~\cite{patchscout,li2024patchfinder}.
For the \texttt{implicit-set}, where ground-truth VFCs are missing, other than Manual Effort@K, we additionally use Precision@K after manually validating the top-K predictions.

\vspace{4px}
\noindent\textbf{Recall@K} measures the proportion of true VFCs that appear within the top-K ranked results.
For a set of queries CVE descriptions $Q$, where each query $q$ has a set of relevant VFCs $R_q$ and retrieved results $K_q$ at rank K.
A high Recall@K indicates that most true VFCs can be found within a manageable number of candidates.
Recall@K is formally defined as:

\begin{equation}
\text{Recall@K} = \frac{1}{|Q|} \sum_{q \in Q} \frac{|R_q \cap K_q|}{|R_q|}
\end{equation}

\vspace{4px}
\noindent\textbf{MRR} (Mean Reciprocal Rank) focuses on the rank of the first correctly identified VFC, providing insight into how quickly users can find a true VFC. For a set of queries $Q$, MRR is calculated as:

\begin{equation}
\text{MRR} = \frac{1}{|Q|} \sum_{q \in Q} \frac{1}{\text{rank}_q}
\end{equation}

where $\text{rank}_q$ is the position of the first true VFC for query $q$. 
MRR is particularly valuable in practical scenarios where developers need to quickly locate at least one true VFC to understand and address the vulnerability.

\vspace{4px}
\noindent\textbf{Manual Effort@K} quantifies the inspection cost.
Prior research in fault localization has shown that practitioners have limited tolerance for examining ranked results, with approximately 74\% of developers expecting to find the target within the top 5 positions, otherwise preferring to resort to manual debugging~\cite{pavneet2016practitioners}. This finding directly applies to VFC identification, where developers similarly need to quickly locate fixing commits among potentially thousands of candidates. Furthermore, this metric has been widely adopted in prior VFC identification studies~\cite{dunlap2024vfcfinder,patchscout,li2024patchfinder, yu2024llm_enhanced_vfc} as it directly measures the practical burden on security practitioners.
It represents the number of candidate commits a developer must inspect to encounter a true VFC, capped at $K$.

In particular, if $\text{rank}_q$ is the first true VFC in the rank list, and the users are asked to check the top K commits predicted to locate the first true VFC, the manual efforts for locating the first true VFC of this CVE are $min(\text{rank}_q,K)$. Lower values correspond to less human effort.
Manual Effort@K is computed as:
 
\begin{equation}
\text{Manual Effort@K} = \frac{1}{|Q|} \sum_{q \in Q} min(\text{rank}_q,K)
\end{equation}

\vspace{4px}
\noindent\textbf{Precision@K} measures the proportion of true VFCs among the top-K ranked results, evaluating the accuracy of the retrieved candidate commits.
Precision@K is calculated as:

\begin{equation}
\text{Precision@K} = \frac{1}{|Q|} \sum_{q \in Q} \frac{|R_q \cap K_q|}{k}
\end{equation}

Different from Manual Effort@K, where we focus on how many commits we need to inspect to meet the first true VFC, Precision@K measures accuracy within the same top-K window 
Thus, while Manual Effort@K asks ``\emph{How long until I see a VFC?},'' Precision@K asks ``\emph{How many of these commits are VFCs?}'' and is especially useful when multiple fixes may exist for a single CVE.






\subsection{Implementation Details}

Each of the backbone models of \method was trained for 3 epochs with a batch size of 32 and a learning rate of 1e-4.
To make the training more effective, we also incorporate a hard negative sampling strategy used by RepLLaMA~\cite{repllama}. 
For each training instance, we select hard negatives that are sampled from the top-ranking results of a hybrid retriever combining BM25~\cite{bm25} and a pre-trained CoCondenser~\cite{gao2021codenser} model.
For the InfoNCE loss function, we set the temperature parameter $\tau$ to 0.01. The training was conducted on a system with four NVIDIA H100 80GB GPUs, while evaluation was performed on a single NVIDIA H100 80GB GPU. 
For the commit message generation, we used the official CCT5 checkpoint~\footnote{https://github.com/Ringbo/CCT5} provided by the original authors, without any additional fine-tuning.

We ran all baselines on our datasets. 
For PatchScout and PatchFinder, we used the publicly available replication packages of PatchFinder and followed their prescribed setup procedures. 
For Prospector, we used its official replication package. However, to align the candidate selection process with our dataset's construction (as described in Section~\ref{sec:dataset}), we modified its default configuration. 
Specifically, we expanded its search window from the default 120-day interval around the CVE's reservation date to a full year, to match the setting of our \method dataset. 
We also increased the maximum number of candidate commits to retrieve from 2,000 to 7,000, since in our \method dataset, the average number of candidate commits for each commit is 6,621.
The execution time limit for each vulnerability remained at 30 minutes. 
All other heuristics and scoring mechanisms remain unchanged from the original implementation.

When conducting the evaluation on the \texttt{implicit-set}, two authors inspected the predicted VFCs produced by different methods.
Specifically, for each predicted VFC, the two authors conducted a comprehensive inspection. 
This process involved reviewing the full code diff, the commit message, and any linked artifacts such as pull requests or issue tracker tickets. 
In addition, they also cross-referenced the commit with the affected and fixed version information provided in the NVD entry and the project's official release notes.
A commit was labeled as a true VFC only if its code change was judged to logically remediate the specific vulnerability described in the NVD. 
Any disagreements between the two authors were resolved through discussion to reach a single final conclusion.

\section{Results}
\label{sec:results}

\begin{table}[t]
\renewcommand{\arraystretch}{0.9}
\centering
\caption{RQ1 results – Recall@K (R@K) and MRR of \method and baselines.}
\small
\begin{tabular}{@{}lcccc@{}}
\toprule
\textit{R@K} & \textbf{PatchScout} & \textbf{PatchFinder} & \textbf{Prospector} & \textbf{\method} \\
\midrule
\rowcolor[HTML]{EFEFEF} 
K=1            & 0.156 & 0.200 & 0.354 & \textbf{0.702} \\
K=2           & 0.232 & 0.275 & 0.451 & \textbf{0.773} \\
\rowcolor[HTML]{EFEFEF} 
K=3           & 0.267 & 0.337 & 0.525 & \textbf{0.811} \\
K=4           & 0.326 & 0.379 & 0.565 & \textbf{0.829} \\
\rowcolor[HTML]{EFEFEF} 
K=5           & 0.372 & 0.413 & 0.597 & \textbf{0.843} \\
K=6           & 0.402 & 0.449 & 0.617 & \textbf{0.852} \\
\rowcolor[HTML]{EFEFEF} 
K=7           & 0.424 & 0.475 & 0.644 & \textbf{0.860} \\
K=8           & 0.478 & 0.499 & 0.658 & \textbf{0.865} \\
\rowcolor[HTML]{EFEFEF} 
K=9           & 0.485 & 0.515 & 0.672 & \textbf{0.868} \\
K=10          & 0.492 & 0.525 & 0.681 & \textbf{0.871} \\
\rowcolor[HTML]{EFEFEF} 
K=20          & 0.540 & 0.615 & 0.747 & \textbf{0.919} \\
K=50          & 0.633 & 0.729 & 0.830 & \textbf{0.948} \\
\rowcolor[HTML]{EFEFEF} 
K=100         & 0.694 & 0.791 & 0.874 & \textbf{0.971} \\ 
\midrule
\textbf{MRR}     & 0.284  & 0.301 & 0.464 & \textbf{0.739}   \\ 
\bottomrule
\end{tabular}
\label{tab:rq1_r}
\end{table}

\begin{table}[t]
\centering
\caption{RQ1 results – Manual Effort@K (ME@K) results of \method and baselines.}
\renewcommand{\arraystretch}{0.9}
\small
\begin{tabular}{@{}lcccc@{}}
\toprule
\textit{ME@K} & \textbf{PatchScout} & \textbf{PatchFinder} & \textbf{Prospector} & \textbf{\method} \\ 
\midrule
\rowcolor[HTML]{EFEFEF} 
K=1           & 1.000 & 1.000 & 1.000 & \textbf{1.000} \\
K=2           & 1.963 & 1.800 & 1.551 & \textbf{1.291} \\
\rowcolor[HTML]{EFEFEF} 
K=3           & 2.754 & 2.524 & 2.057 & \textbf{1.549} \\
K=4           & 3.441 & 3.186 & 2.552 & \textbf{1.763} \\
\rowcolor[HTML]{EFEFEF} 
K=5           & 4.103 & 3.806 & 3.054 & \textbf{1.945} \\
K=6           & 4.703 & 4.392 & 3.550 & \textbf{2.122} \\
\rowcolor[HTML]{EFEFEF} 
K=7           & 5.308 & 4.942 & 4.053 & \textbf{2.281} \\
K=8           & 5.901 & 5.466 & 4.551 & \textbf{2.439} \\
\rowcolor[HTML]{EFEFEF} 
K=9           & 6.497 & 5.966 & 5.058 & \textbf{2.585} \\
K=10          & 7.002 & 6.450 & 5.556 & \textbf{2.734} \\
\rowcolor[HTML]{EFEFEF} 
K=20          & 12.481 & 10.774 & 9.757 & \textbf{3.941} \\
K=50          & 23.915 & 20.108 & 18.231 & \textbf{6.179} \\
\rowcolor[HTML]{EFEFEF} 
K=100         & 37.428 & 31.936 & 29.120 & \textbf{8.262} \\ \bottomrule
\end{tabular}
\label{tab:rq1_me}
\end{table}

\subsection{RQ1: Comparison with Baselines}

As shown in Table~\ref{tab:rq1_r} and Table~\ref{tab:rq1_me}, \method demonstrates a clear and significant performance advantage over all baseline methods across every evaluation metric.
In particular, \method achieves an MRR of 0.739, a substantial improvement over the baselines: 146\% higher than PatchFinder, 160\% higher than PatchScout, and 59.3\% higher than Prospector. 
This superiority is also reflected in its recall; at K=1, \method's recall of 0.702 nearly doubles Prospector's 0.354 and is significantly higher than the other learning-based methods.
We also observe that \method requires significantly less manual effort compared to the baselines.
When $K = 1$, all methods have a Manual Efforts value of 1, as this is inherently bounded by $K$, regardless of whether the true VFC is found.
As $K$ increases, however, \method consistently demonstrates much lower effort than the other three baselines.
For example, at $K = 100$, \method requires only 8.262 manual efforts, which is less than one-third of the effort required by the other methods.

\begin{figure}[h]
    \centering
    \includegraphics[width=0.7\columnwidth]{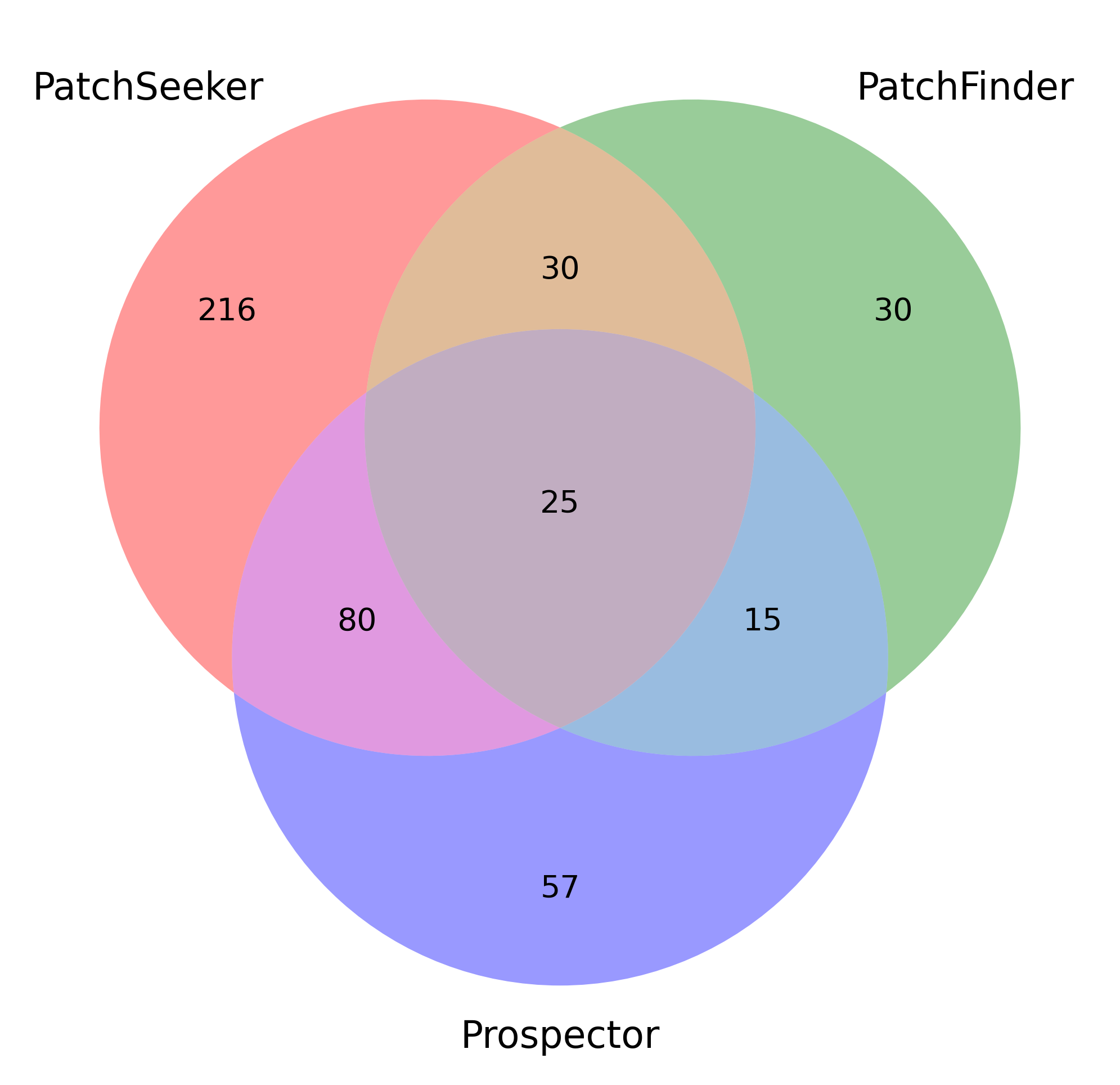}
    \caption{Venn diagram of CVEs that \method, PatchFinder, and Prospector correctly identified VFCs on the first prediction.}
    \label{fig:venn_rq1}
\end{figure}

Figure~\ref{fig:venn_rq1} illustrates the complementary strengths of the different approaches.
While there is a notable overlap in the VFCs identified, \method successfully finds 216 VFCs that both PatchFinder and Prospector miss entirely, highlighting its unique capability to handle cases that are too complex for the other methods.
Furthermore, \method encompasses the majority of VFCs found by the baselines.
The results reveal that \method's LLM-based semantic understanding approach outperforms both prior learning methods and heuristic approaches. 
While Prospector performs better than the other learning-based baseline, PatchFinder, it still falls short of \method's performance.

\find{
\textbf{Answer to RQ1:} \method outperforms the prior best-performing method Prospector by 27.9\%, 59.3\%, and 50.8\%, in terms of Recall@10, MRR, and Manual Efforts@10.
In addition, \method can rank more true VFCs in the first prediction than the baselines.
}

\subsection{RQ2: Ablation Study}

\begin{figure*}[t]
  \centering
  \begin{subfigure}[t]{0.42\textwidth}
    \centering
    \includegraphics[width=\linewidth]{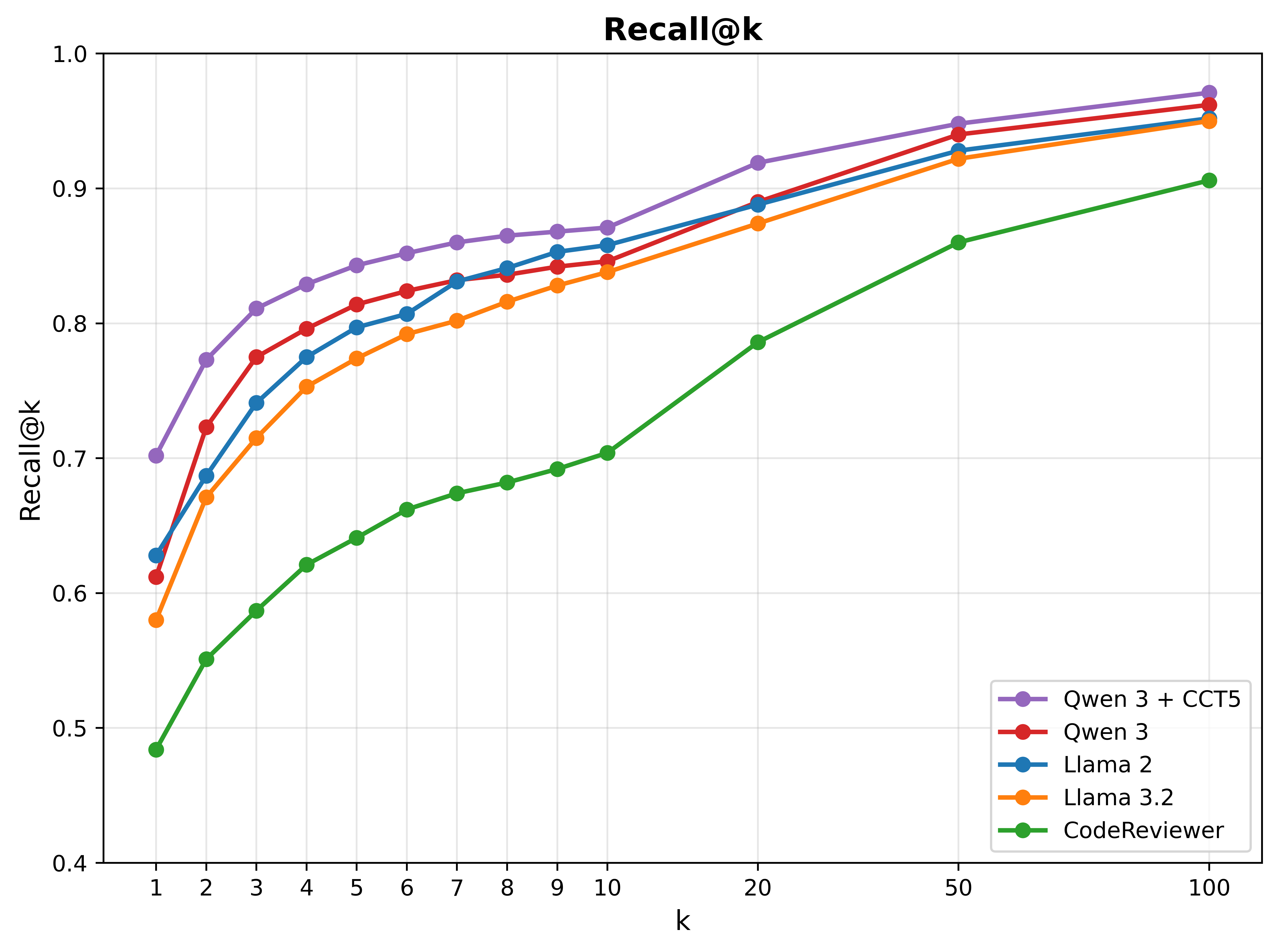}
  \end{subfigure}
  \hfill
  \centering
  \begin{subfigure}[t]{0.42\textwidth}
    \centering
    \includegraphics[width=\linewidth]{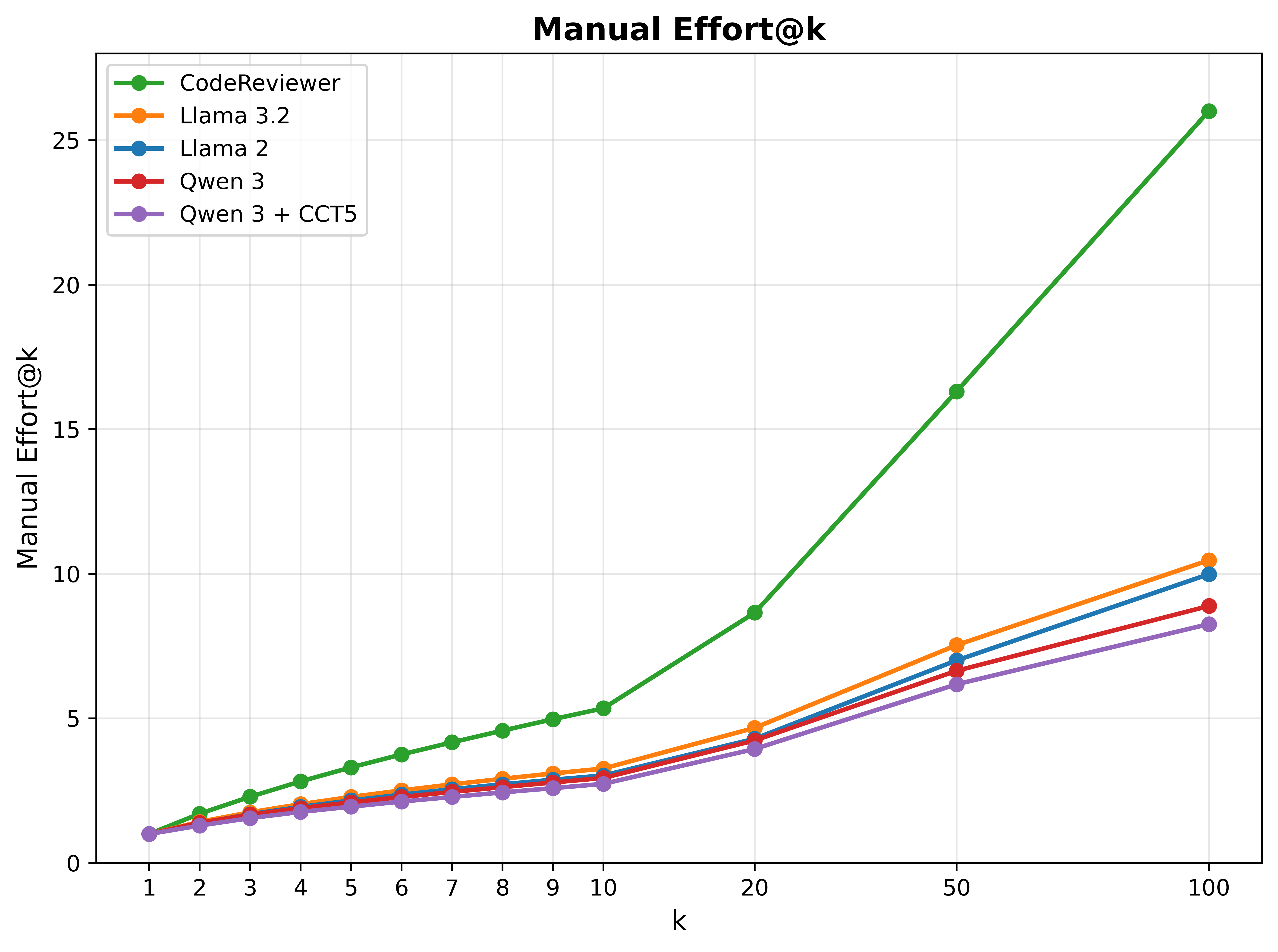}
  \end{subfigure}

  \captionsetup{skip=0pt} 
  \caption{Results of RQ2: Ablation study of \method}
  \label{fig:rq2}
\end{figure*}

\noindent\textbf{Impact of CCT5-Generated Commit Messages:} 
Figure~\ref{fig:rq2} shows the results of with and without CCT5-generated commit messages (Qwen 3 vs. Qwen 3 + CCT5). 
The result reveals that by generating new messages for the 27.3\% of commits with short and uninformative original messages, CCT5 provides consistent improvements across different metric results. 
To be more specific, the generated messages have a clear impact at the top of the ranking, improving Recall@1 from 0.612 to 0.702 (+14.5\%). 
This trend continues for larger result sets, where Recall@50 increases from 0.919 to 0.948 (+3.2\%), and Manual Efforts@50 decreases from 0.817 to 0.825 (+1\%), suggesting that CCT5 helps identify and properly rank VFCs that might have less informative original commit messages. 
This demonstrates that CCT5's ability to generate descriptive commit messages from code diffs adds valuable semantic information that enhances \method's matching capabilities, particularly for commits with vague or non-descriptive original messages.




\vspace{4px}
\noindent\textbf{Impact of Backbone Model Selection:} The choice of backbone LLM significantly influences \method's performance, with substantial variations across models.
LLaMA 2 achieves the highest MRR of 0.746 and Recall@50 of 0.928 but shows lower Recall@1 at 0.628, suggesting it excels at eventually finding VFCs but may not rank the most obvious ones at the top positions. 
Qwen 3 demonstrates the most balanced performance with strong Recall@1 of 0.612 and the lowest Manual Efforts@50 of 0.817, making it particularly effective for practical use cases where both immediate identification and overall ranking quality matter. 
CodeReviewer, despite being specifically pre-trained for code understanding, surprisingly underperforms with the lowest scores across all metrics (MRR: 0.56, Recall@50: 0.86), suggesting that general-purpose LLMs may better capture the semantic relationships between vulnerability descriptions and commits. 
LLaMA 3.2 shows intermediate performance across all metrics.

\find{
\textbf{Answer to RQ2:} CCT5-generated commit messages improve \method's performance, boosting Recall@1 by 14.7\%, while the backbone model selection is also critical, as the Recall@1 performance varies by up to 45\% between the best-performing (Qwen 3) and worst-performing (CodeReviewer) models.}

\subsection{RQ3: Performance in Recent CVEs}

\begin{table}[ht]
\renewcommand{\arraystretch}{0.9}
\centering
\caption{RQ3 results comparing \method, with baselines on the \texttt{explicit-set}. Performance is measured using Recall@K (R@K) and Manual Efforts@K (ME@K).}
\centering
\small
\begin{tabular}{@{}lcccccc@{}}
\toprule
             & \multicolumn{2}{c}{\textbf{PatchFinder}} & \multicolumn{2}{c}{\textbf{Prospector}} & \multicolumn{2}{c}{\textbf{\method}} \\ \midrule
\textbf{MRR} & \multicolumn{2}{c}{0.282}                & \multicolumn{2}{c}{0.451}               & \multicolumn{2}{c}{\textbf{0.658}}                  \\ \midrule
\textbf{K}   & \textbf{R@K}       & \textbf{ME@K}       & \textbf{R@K}       & \textbf{ME@K}      & \textbf{R@K}             & \textbf{ME@K}            \\ \midrule
\rowcolor[HTML]{EFEFEF} 
1            & 0.19               & 1                   & 0.321              & 1                  & \textbf{0.583}           & \textbf{1}               \\
2            & 0.259              & 1.652               & 0.416              & 1.288              & \textbf{0.647}           & \textbf{1.082}           \\
\rowcolor[HTML]{EFEFEF} 
3            & 0.307              & 2.425               & 0.482              & 1.812              & \textbf{0.672}           & \textbf{1.403}           \\
4            & 0.354              & 3.018               & 0.519              & 2.253              & \textbf{0.689}           & \textbf{1.698}           \\
\rowcolor[HTML]{EFEFEF} 
5            & 0.402              & 3.561               & 0.551              & 2.664              & \textbf{0.702}           & \textbf{1.989}           \\
6            & 0.424              & 4.088               & 0.566              & 3.051              & \textbf{0.724}           & \textbf{2.274}           \\
\rowcolor[HTML]{EFEFEF} 
7            & 0.446              & 4.591               & 0.59               & 3.425              & \textbf{0.743}           & \textbf{2.541}           \\
8            & 0.467              & 5.087               & 0.604              & 3.791              & \textbf{0.760}           & \textbf{2.801}           \\
\rowcolor[HTML]{EFEFEF} 
9            & 0.489              & 5.582               & 0.615              & 4.154              & \textbf{0.773}           & \textbf{3.058}           \\
10           & 0.511              & 6.215               & 0.624              & 4.609              & \textbf{0.786}           & \textbf{3.387}           \\
\rowcolor[HTML]{EFEFEF} 
20           & 0.590              & 10.992              & 0.689              & 8.107              & \textbf{0.810}           & \textbf{5.923}           \\
50           & 0.692              & 22.463              & 0.773              & 16.488             & \textbf{0.845}           & \textbf{11.996}          \\
\rowcolor[HTML]{EFEFEF} 
100          & 0.784              & 37.641              & 0.825              & 27.579             & \textbf{0.893}           & \textbf{20.087}          \\ \bottomrule
\end{tabular}%
\label{tab:rq3-explicit}
\end{table}

\begin{table}[ht]
\renewcommand{\arraystretch}{0.9}
\centering
\small 
\caption{RQ3 results comparing \method, with baselines on the \texttt{implicit-set}. Performance is measured using Precision@K (P@K) and Manual Efforts@K (ME@K).}
\begin{tabular}{@{}lcccccc@{}}
\toprule
\multirow{2}{*}{\textbf{K}} & \multicolumn{2}{c}{\textbf{PatchFinder}}         & \multicolumn{2}{c}{\textbf{Prospector}}          & \multicolumn{2}{c}{\textbf{PatchSeeker}}           \\
\cmidrule{2-7}
           & \textbf{P@k} & \multicolumn{1}{l}{\textbf{ME@k}} & \textbf{P@k} & \multicolumn{1}{l}{\textbf{ME@k}} & \textbf{P@k}   & \multicolumn{1}{l}{\textbf{ME@k}} \\ \midrule
           \rowcolor[HTML]{EFEFEF} 
1          & 0.13         & 1                                 & 0.21         & 1                                 & \textbf{0.37}  & \textbf{1}                        \\
2          & 0.152        & 1.88                              & 0.303        & 1.511                             & \textbf{0.521} & \textbf{1.153}                    \\
\rowcolor[HTML]{EFEFEF} 
3          & 0.237        & 2.73                              & 0.390        & 2.184                             & \textbf{0.567} & \textbf{1.591}                    \\
4          & 0.263        & 3.5                               & 0.428        & 2.716                             & \textbf{0.598} & \textbf{1.974}                    \\
\rowcolor[HTML]{EFEFEF} 
5          & 0.281        & 4.24                              & 0.447        & 3.209                             & \textbf{0.634} & \textbf{2.331}                    \\
6          & 0.326        & 4.98                              & 0.485        & 3.684                             & \textbf{0.667} & \textbf{2.668}                    \\
\rowcolor[HTML]{EFEFEF} 
7          & 0.325        & 5.66                              & 0.492        & 4.137                             & \textbf{0.681} & \textbf{2.989}                    \\
8          & 0.347        & 6.34                              & 0.501        & 4.576                             & \textbf{0.692} & \textbf{3.303}                    \\
\rowcolor[HTML]{EFEFEF} 
9          & 0.354        & 7                                 & 0.510        & 5.021                             & \textbf{0.703} & \textbf{3.614}                    \\
10         & 0.361        & 7.65                              & 0.519        & 5.573                             & \textbf{0.710} & \textbf{4.002}                    \\ \bottomrule
\end{tabular}%
\label{tab:rq3-implicit}
\end{table}


\vspace{2px}
\noindent\textbf{Performance on CVEs from the \texttt{explicit-set}.}
Table~\ref{tab:rq3-explicit} reports the effectiveness of \method and the two best-performing baseline methods based on their results in RQ1, i.e., PatchFinder and Prospector, on the \texttt{explicit-set}.
\method exhibits a clear lead across all metrics.  
It achieves an MRR of 0.658, more than 133\% higher than PatchFinder and 46\% higher than Prospector. 
Looking at the strictest cut-off, Recall@1 reaches 0.583, meaning that in 58.3\% of the cases \method ranks the correct VFC in the first position.
This rate is \(\approx\!3.1\times\) better than PatchFinder and 81\% better than Prospector.
The advantage persists when we inspect a larger result window: at \(k=10\) \method returns at least one correct VFC for 78.6\% of the CVEs (Recall@10), outperforming PatchFinder by 53.8\% and Prospector by 27\%. 
Even looking at even large window, i.e., \(k=100\), \method still shows an advantage (Recall@100~\(=0.893\) vs.\ 0.784/0.825).
Manual Effort@K follows the same trend, confirming that \method not only finds more VFCs but also requires less manual effort to locate the first true VFC.

\vspace{2px}
\noindent\textbf{Performance on CVEs from the \texttt{implicit-set}.}
Table~\ref{tab:rq3-implicit} reports the effectiveness of \method, and two baselines, PatchFinder and Prospector, on the \texttt{implicit-set}.
CVEs from the \texttt{implicit-set} are considerably harder, given that there is no explicit ground-truth VFCs, and possibly some CVEs may not yet be patched, resembling real-world scenarios.  
Nevertheless, \method remains the strongest model.  
At \(k=1\), it obtains a Precision of 0.37, nearly \({3\times}\) PatchFinder's 0.13 and \(76\,\%\) above Prospector's 0.21. 
For k = 10, \method's Precision@10 climbs to 0.71. 
This represents a \(97\,\%\) improvement over PatchFinder and a \(37\,\%\) margin over Prospector.  
This suggests that \method would find a true VFC in roughly seven of every ten commits developers inspect, substantially reducing the manual effort compared to current baselines.

\textit{Why do implicit CVEs remain challenging?}
The accuracy drop from explicit to implicit CVEs is observable for every model we tested. 
There are two potential justifications behind this. 
First, implicit CVEs are more likely to have incomplete reports in the NVD that lack detailed information and supporting evidence compared to explicit CVEs. 
These reports may be missing key details about the vulnerability, making it harder for retrieval models to find matching commits.  
Second, the ground truth itself is uncertain: some implicit CVEs might describe bugs that were never fixed in the main code, or were only fixed in private copies. 
This means even a perfect model would seem to fail when we test it. 
These observations match the findings of the large-scale repository study by Nguyen et al. \cite{nguyen2025mappingnvdrecordsvfcs}: only \(\sim\!8\,\%\) of \textsc{NVD} entries contain Git references, and for the vast majority of records the VFCs must be \emph{mined} from surrounding evidence.
Therefore, our results underline two complementary research directions: (1) improving retrieval models to cope with sparse descriptions and (2) expanding benchmark datasets beyond Git-referenced CVEs to mitigate evaluation bias.

\find{
\textbf{Answer to RQ3:} \method retains its superiority on the recent CVEs, achieving up to \({133\,\%}\) higher MRR and \({53.8\,\%}\) higher Recall@10 than PatchFinder on the \texttt{emplicit-set}, while almost doubling Precision@10 on the \texttt{implicit-set}.
}

\section{Discussion}
\label{sec:discussion}


\subsection{Qualitative Analysis}
In this section, we conducted qualitative analysis to better understand the specific advantages and limitations of \method.
Our analysis focuses on the results of the main benchmark dataset.

\noindent\textbf{Success cases.}
We manually investigated a set of randomly sampled 10 CVEs where \method successfully identified the true VFCs in its top-ranked result, but both PatchFinder and Prospector failed to do so. 
Our goal was to identify patterns in these cases to categorize the specific challenges that the baselines could not overcome.
Our analysis revealed a common pattern across all these failures: the baselines lack the deep semantic understanding required to connect a vulnerability description with a commit message when they do not share obvious keywords.

The core challenge in all 10 cases was the need to understand the conceptual relationship between different sets of technical terms. 
For example, in the case of CVE-2023-3071, the NVD description mentions a ``Cross-site Scripting (XSS)'' vulnerability. The corresponding VFC message, however, reads ``Sanitize picklist values,'' which uses implementation-specific language to describe the fix. 
While conceptually linked, these phrases have no overlapping keywords. 
\method successfully identified this connection, while the baselines failed for distinct reasons tied to their architectures.
On one hand, PatchFinder relies on a small model, CodeReviewer, which is not powerful enough to grasp these abstract semantic relationships.
On the other hand, Prospector is explicitly programmed to look for predefined keywords and patterns. It has no mechanism for understanding synonyms or abstract concepts. 



\begin{lstlisting}[caption=A failure case of \method., label={lst:failure_case}, float=t]
(1) CVE Description of CVE-2023-25667:
"TensorFlow is an open source platform for machine learning. Prior to versions 2.12.0 and 2.11.1, integer overflow occurs when $ 2^{31} \le num\_frames * height * width * channels < 2^{32} $, for example Full HD screencast of at least 346 frames. A fix is included in TensorFlow version 2.12.0 and version 2.11.1."

(2) Commit message of the true VFC, which PatchSeeker ranks at #216:
"Fix integer overflow for multiframe GIFs."

(3) Commit message of a false VFC, which PatchSeeker ranks at #1:
"Fix histogram overflow vulnerability. Changed order of cast and subtraction to avoid overflows."
\end{lstlisting}

\vspace{2px}
\noindent\textbf{Failure cases.}
Listing~\ref{lst:failure_case} shows a case of CVE-2023-25667, which involves an integer overflow vulnerability in TensorFlow. 
We found that \method ranked the true VFC at a distant \#216 position.
Instead, it incorrectly promoted a different commit to the top rank.
Our analysis suggests two likely reasons for this failure. 
Firstly, the top-ranked incorrect commit's message contained strong semantic cues that aligned closely with the CVE description. The repeated use of the word "overflow" and the mention of "subtraction" likely created a high similarity score with the NVD text, which describes an "integer overflow" and includes a mathematical formula. 
Secondly, we found that \method's failure was rooted in its tokenization strategy. The Qwen~3 tokenizer splits the word ``multiframe'' into the parts ``mult'' and ``iframe''. This unusual split the critical ``frame'' signal from the model, preventing it from making a direct connection to the ``num\_frames'' keyword in the CVE description.

In contrast, PatchFinder succeeded in this case precisely because of a difference in tokenization. Its model, CodeReviewer, utilizes the tokenizer from RoBERTa, which correctly splits the compound word ``multiframe'' into the more intuitive subwords ``multi'' and ``frame''. This seemingly minor difference is critical. By exposing the semantically correct ``frame'' token, the model could establish a direct lexical link to ``num\_frames'' in the CVE description, allowing PatchFinder to correctly identify and prioritize the true VFC.

The failure to understand "multiframe" highlights a specific limitation where a more simplistic, keyword-aware approach, aided by favorable tokenization, can outperform a purely semantic one. 
Future work should explore hybrid models or tokenization enrichment strategies to be more robust against domain-specific jargon and compound words.

\subsection{Threats to Validity}
\noindent\textbf{Threats to Internal Validity:} We leverage the MoreFixes~\cite{morefixes} dataset to build our benchmark dataset. 
Although Akhoundali el al.~\cite{morefixes} already tried to reduce the noise in the MoreFixes dataset as much as possible by evaluating it in a manual review and selecting a reasonable score threshold, the resulting dataset might still include some irrelevant commits. 
We inherit the threats from their method.
In addition, the CVE and CWE information can be inaccurately labeled in NVD; we therefore inherit this threat.

\vspace{2px}
\noindent\textbf{Threats to Construct Validity:} 
The first potential threat involves the suitability of our evaluation metrics.
To mitigate this threat, we follow prior studies~\cite{patchscout,li2024patchfinder,wang2022vcmatch} on VFC retrieval to use Recall@K, MRR, and Manual Effort@K.
For the \texttt{implicit-set}, we additionally report Precision@K to capture precision‑oriented performance.
As these metrics are widely adopted, we deem such a threat to be limited.
A second threat arises from the subjectivity inherent in the manual verification on the \texttt{implicit-set}. 
We addressed this by having two authors perform the labeling process, thereby enhancing the reliability of our labels through cross-validation.

\vspace{2px}
\noindent\textbf{Threats to External Validity:} 
Our benchmark covers 5,000 CVEs drawn from 2,070 repositories spanning multiple programming languages, and we further evaluate on a set of recent CVEs.
While the breadth of our analysis does not completely eliminate the risk of dataset-specific conclusions, it substantially reduces this threat. 
Consequently, we regard the remaining external threats as minor.

\section{Related Work}
\label{sec:related}
In this section, we briefly review three research directions that are most relevant to our work.

\vspace{2px}
\noindent\textbf{Mapping NVD records to their VFCs.} Most of the early studies directly consider the Git-related references in NVD as the VFCs~\cite{bigvul,bhandari2021cvefixes,ding2024primevul,morefixes, zhou2019devign}.
For instance, BigVul~\cite{bigvul} uses the BeautifulSoup tool to web-scrape all the Git-related references from the CVE database. 
CrossVul\cite{nikitopoulos2021crossvul} and CVEFixes\cite{bhandari2021cvefixes} follow a similar technique, which is collecting GitHub commits and GitHub pull requests from the references sections of NVD. Additionally, some studies extend their search beyond the NVD by looking at external security databases. 
One example is DiverseVul~\cite{chen2023diversevul}, which retrieves VFCs from the Snyk~\cite{snyk} and Bugzilla~\cite {bugzilla} security databases. 
However, these traditional approaches are limited by the existing explicit links in these security databases.
Later, with the emergence of learning methods, various works have tried to mine more VFCs from CVEs that the traditional approaches fail. 
VFCFinder~\cite{dunlap2024vfcfinder} uses CodeBERT~\cite{feng2020codebert} models with five intuitive features to identify the VFCs.

\vspace{2px}
\noindent\textbf{Automated commit message generation.} Automated commit message generation has been a focus of significant research to reduce developer burden and improve software maintenance. 
Existing approaches can be categorized into three main types: rule-based, learning-based, and pre-trained models.
Rule-based approaches generate messages using predefined templates to summarize code changes such as DeltaDoc~\cite{buse2010deltadoc}and ChangeScribe~\cite{cortes2014ChangeScribe}.
Learning-based approaches frame the problem as a translation task, using neural machine translation to convert code diffs into messages. Notable works include CmtGen~\cite{jiang2017cmtgen}, NNGen~\cite{liu2018nngen}, and CODISUM~\cite{xu2019codisum}.
Recent work has also explored pre-trained models for diff encoding, such as CCRep~\cite{liu2023ccrep}, FIRA~\cite{fira}, and CCT5~\cite{lin2023cct5}, which leverages CodeBERT~\cite{feng2020codebert} and CodeT5~\cite{wang2021codet5} respectively for improved representations.
Despite these advances, significant challenges remain, including handling low-quality data and processing large diffs beyond typical 100-200 token limits~\cite{zhang2024commit_msg}.

\vspace{2px}
\noindent\textbf{LLM-enhanced software vulnerability analysis.} LLMs have brought considerable breakthroughs in software engineering research~\cite{zhang2025revisiting}, including software vulnerability analysis.
For instance, some studies have investigated LLMs' capability in detecting software vulnerabilities~\cite{zhou2024large,zhang2025benchmarking,ding2024primevul}.
These studies typically formulate the task as a classification problem, fine-tuning LLMs with a binary classification loss, or they utilize the inherent generative capabilities of LLMs without fine-tuning. 
Nevertheless, leveraging LLMs for retrieval-based analysis represents a significant, yet underexplored, research avenue. 
As an information retrieval task characterized by a large pool of candidate commits, mapping NVD records to their VFCs is not suitable for the above classification or generation settings.
Our work fills this gap by systematically exploring the performance of LLMs on this task.

\section{Conclusion and Future Work}
\label{sec:conclusion}
In this work, we have proposed a new method \method for mapping NVD records to their VFCs.
\method leverages an LLM to embed both CVE descriptions and candidate commit messages in a shared semantic space.
When a raw commit message is overly brief, \method augments it by generating a richer message from the code changes.
We provide a benchmark that consists of 5,000 CVEs and 5,539 true VFCs from 2,070 projects.
On this benchmark, \method outperforms the prior best-performing approach, Prospector, by 27.9\% in terms of Recall@10 and 59.3\% in terms of MRR.
To assess generalizability and further demonstrate the effectiveness of \method, we collect CVEs published in the recent 6 months and categorize them into two subsets, i.e., \texttt{explicit-set} and \texttt{implicit-set}.
Our method \method continuously demonstrates better performance, i.e., it outperforms Prospector by 27\% in terms of Recall@10 on the \texttt{explicit-set}, 
and 37\% in terms of Precision@10 on the \texttt{implicit-set}.

In the future, we plan to deepen our understanding of commit message generation.
Specifically, we would like to investigate which linguistic cues most accurately convey vulnerability‑fixing intent and how enriched messages can further boost CVE‑to‑VFC matching accuracy.
In addition, we want to explore more ways to leverage the augmented commit messages instead of simply concatenating them with the raw commit messages.
Our replication package is at: \url{https://anonymous.4open.science/r/PatchSeeker-ECFD/}.

\newpage
\balance
\bibliographystyle{ACM-Reference-Format}
\bibliography{main}

\end{document}